\input amstex \magnification=\magstep 1 \baselineskip=24pt
\def\footnote{\plainfootnote}
\magnification=\magstep1 \baselineskip=24pt


\centerline {\bf Some Special Cases of Khintchine's Conjectures in Statistical Mechanics:}
\centerline{\bf Approximate Ergodicity of the Auto-Correlation Functions }
\centerline{\bf of an Assembly of Linearly Coupled Oscillators}
\centerline   {by Joseph F.\ Johnson}
\centerline   {Dept.\ of Mathematics and Statistics, Villanova University}

\centerline {\bf Abstract}
{\narrower\smallskip\baselineskip=12pt
\noindent
In order to estimate the theoretical auto-correlation function of a time series
from the sample auto-correlation function of one of its realisations, it is
usually assumed without justification that the time series is ergodic. 
In 1943, Khintchine made some visionary conjectures about dynamical systems with
large numbers of degrees of freedom which would justify, even in the absence of
ergodicity, approximately the same conclusions. 
We prove Khintchine's conjectures in some special cases of a linearly coupled
assembly of harmonic oscillators. 
\smallskip}


\noindent\bf RESUMEN\rm \ 

{\narrower\smallskip\baselineskip=12pt
Para emplear el correlograma de los valores muestrales de un proceso estoc\'astico para estimar su funci\'on te\'orica de autocorrelaci\'on,
por regla general se asume, sin justificaci\'on, que el proceso 
es erg\'odico.  Pero en 1943, Khintchine conjetur\'o proposiciones
de gran importancia en este asunto, que justificar\'\i an una aproximaci\'on a las 
mismas estimaciones a\'un sin la ergodicidad del sistema.
Mostraremos casos particulares de las conjeturas de Khintchine
para asambleas de osciladores lineales.
\smallskip}

\input amstex \magnification=\magstep 1 \baselineskip=24pt
\loadmsbm

\centerline {\bf Preface}

A novel way to justify the 
use, in Statistical Mechanics, of the equality of time averages with phase averages 
was envisioned by Khintchine in 1943, but he could only prove 
special cases [9], [10]. 
He suggested that for quite general, non-ergodic dynamical systems, 
a kind of approximate ergodicity for a restricted class of observables should 
arise when the number of degrees of freedom is sufficiently large.
([9], pp.\ 62--63).

\vskip -8pt
In the study of Brownian motion by Ford--Kac--Mazur 
in 1965 [2] in terms of a Hamiltonian 
heat bath  they carried out the Gibbs program in detail for a concrete, linear, 
Hamiltonian model.  

\vskip -8pt
The task of this paper is to recast the model of Brownian 
motion to show that one may assume a determined initial condition and derive a 
stochastic process in the thermodynamic limit without assuming any initial probability 
distribution.  It extends Khintchine's vision to an essentially new case (earlier extensions by Ruelle [13] and Lanford [8] assumed weak, short-range interactions).  It is already known 
that the results of [2] remain true for many different choices of an initial 
distribution (see Kim [11] for a survey).

\vskip -8pt
The systems we study are linear Hamiltonian systems and are very far from being 
ergodic.  But as Khintchine foresaw, a kind of approximate ergodicity holds good 
for some measurable functions (having particular physical significance) when the number of degrees of freedom is very 
large.  In our case, the measurable functions we study are the auto-correlations 
of the time-evolution of the momentum coordinate of one of the particles.

This paper falls into two halves which are almost separate: the first half is a survey
of the problem and does not pretend to any originality, given at a seminar at the 
Univ. of Havana in 2011.  The second half presents the technical details which
were left out of the seminar and are original.

\centerline {\bf Introduction to the Role of Ergodicity in the Theory of Time Series}

\noindent Time Series: Two Contradictory Definitions

\vskip -8pt
The notion of time series has two definitions which although
related cause confusion to the student.  The first meaning is
that a time series is a series of data distributed in time.
If, for example, $M$ is a dynamical system with Hamiltonian
$H(p,q) = \frac12(|p|^2+|q|^2)$, where $p$ and $q$ are $n+1$-dimensional
vectors,  i.e., a collection of $n+1$ uncoupled harmonic oscillators, 
then since $\dot p=-\frac{\partial H}{\partial q}$ and 
$ \dot q= \frac{\partial H}{\partial p}$, $p(t)$ is a time 
series if $p(0)$ and $q(0)$ are given.  Another example
is the (infinite) sequence 
$$011110111111111110010111111111101111111 \dots $$
of tosses of a fair coin (there would not be any contradiction of either the
laws of Physics or the laws of Probability if all future tosses resulted in 1).
 These are both examples
of deterministic data since even the coin toss is a function
of time in the usual, deterministic, sense of the word 
\it function\rm.

\vskip -8pt
The second meaning of \it time series\rm,
is that it is a sequence of random variables.  This sense is
also called a \it stochastic process\rm.  Two examples which
are related to the previous examples are:

\noindent Coin Toss: $X_n$ a sequence of independent, identically 
distributed random variables taking the values 1 and -1 with equal probabilities.
In fact, the space of all possible sequences of results of 
a coin toss, i.e., the space of all possible sequences of binary 
digits, can be mapped to the unit interval $[0,1]\subset\Bbb R$ 
by regarding each sequence $x_n$ as the binary expansion of the real number
$$\sum_0^\infty {(1+x_n)\over2^{n+2}}.$$
This map is an equivalence of probability spaces (it is one to one except on a set of 
measure zero), between
the space of all possible sequences of tosses and the unit interval with Lesbegue measure. 

\vskip -8pt
\noindent Dynamical System: put some (any) probability distribution
on the set of initial conditions $\{ p(0),q(0) \}= \Bbb R^{2(n+1)}$,
for example, the Maxwell--Boltzmann distribution 
$$ (2\pi kT)^{n+1} \det A ^{-\frac12} e^{-H(p(0),q(0))\over kT  }$$
where $k$ is Boltzmann's constant and $T$ is the absolute temperature 
in Kelvin.  (For later generalisations, we here allow the possibility of 
a linear coupling given by a matrix $A$.) Then $p(0)$ is now a random variable and so is $p(t)$ for 
any $t$, so $\{p_o(t)   \}_t $ is a continuous series of random variables
and hence a time series in the sense of a stochastic process.

\vskip -8pt
From the standpoint of the rest of statistics, time series are odd
and difficult because we have to regard the time series in the second
sense as the population and the time series in the first sense as 
one sample taken from the population.  For example, if the probability
space of the random variables $X_n$ from the example of the coin toss
is taken to be the unit interval $[0,1]$ with Lebesgue measure $dx$, then
for any fixed $\alpha \in [0,1]$ we get a time series of data, that is, 
a time series in the first sense, given by $X_n(\alpha)$.
(Every sample point has probability zero, so this is an example of 
a probability space where probability zero does not mean `impossible'.)
Unlike the examples of data in first-year statistics courses, we can never draw more than
one sample point since, e.g., we cannot go back to the year 2000 and 
`try again'.  

\vskip -8pt
Other terms used are, e.g., that
$\{p_o(t)   \}_t $ is an ensemble of time series (in the first sense),
and that given a particular value $(p(0),q(0))$ for the initial conditions,
then $p_o$ is a well-defined function of $t$ called \it a realisation \rm of
the time series.  
These two senses are intimately related, the confusion in 
terminology serves a useful purpose, and it is not going to be reformed any time
soon.

\noindent Statistics of Time Series.

The usual descriptive statistics from first-year statistics courses are less
useful here.  The average is misleading if the time series possess a trend.
Trends and cycles are more important than the average or dispersion.  The most
important descriptive statistic of a time series is  
its auto-correlation function or correlogram.  
(I wish to emphasise here that it is a descriptive statistic: it has no more 
probabilistic significance than did the average or standard deviation.)
Given a series of data $f(t)$
it measures the average influence of $f(t)$ on $f(t+\tau)$ and is given by 
$$\varphi(\tau) = \lim_{T\rightarrow\infty}\frac1T\int _0^T f(t)f(t+\tau) dt .$$
This is called the sample auto-correlation function when it is necessary to 
distinguish it from a related notion which does not use the time average
but uses the whole theoretical model (population) instead of only one realisation
of the process (the data), 
the phase or population or model or `theoretical' auto-correlation function:
$$R(\tau) = E(X_n X_{n+\tau}).$$
Here, $E$, the expectation, is taken over the probability space, which is sometimes
the phase space of a dynamical system, so it can be regarded as a phase average
as opposed to the sample auto-correlation function, which was a time average 
of actual given data.  (For simplicity we assume, from now on, that all random variables are \it centred\rm, i.e., have zero expectation.)

\vskip -8pt
\noindent Examples.

\vskip -8pt
\noindent 1. Given a data stream $p(t) = \sum a_n \cos nt + \sum b_n \sin nt$, we obtain
$$ \varphi(\tau) = \sum {|a_n|^2 + |b_n|^2 \over 2} \cos n\tau,$$
which is an even function.

\noindent 2. If the process $X(t) = p(t)$ `with random phases', i.e., if each $p(t)$ is turned into a random variable by introducing random phase shifts in its terms, then $R(\tau)$ is the same thing.
(The phases can be uniformly distributed or Gaussian, it makes no difference.)

\noindent 3. Coin toss: $$E(X_nX_{m}) = 0 \text{  if } n\neq m$$
by independence.  $R(\tau) = \delta(\tau)$, a spike.

\noindent 4. Data from a coin toss: $\varphi_\alpha(\tau)$ could be any positive definite
function (that takes the value 1 at the origin).  For example, one realisation could have 
a periodic correlogram, a sawtooth alternating between 1 and -1, and another set of tosses (all heads) could yield
a constant function.  Neither of these is very close to the theoretical auto-correlation
function calculated above, but a `normal' realisation will have a correlogram close to a spike.

Since we only have one sample point from the population, the problem is how 
to infer $R(\tau)$ from $\varphi(\tau)$?  This is the topic of this paper. 
The answer has usually been taken to be
the concept of \it ergodicity\rm, a concept borrowed from Statistical Mechanics to 
which we will turn in the next section.  The reader should be warned that within the discipline of time series, the 
term `large sample theory' has been perverted from its meaning in the rest of statistics, since
here its original meaning is largely irrelevant (and we will not use it in this lecture).
Within advanced time series texts, it means the theory of one sample point which has a lot of data in it. (See the careful discussion in Fuller [3], pp.\ 308ff.)

\noindent L\'evy's philosophy.

L\'evy pioneered the method of replacing the study of a stochastic process by
a study of its auto-correlation function (sometimes called the auto-covariance
function or sometimes normalised in a certain fashion).
His philosophy [12] was that for a wide class of stochastic processes, all important
properties can be seen in the theoretical auto-correlation function of the process.

\noindent Statistical Mechanics

If $M$ is a dynamical system, as above, and one supposes that even non-engineering data sets such as those of climate change or coin tosses are indeed the results of an immensely complicated dynamical system with an astronomical number of degrees of freedom, then a measurable function $f$ on $M$ is called an observable.  But in fact $f$ itself is not observable. A measurement of $f$ is always \it macroscopic\rm, it is always the result of letting some part of the system come into contact with a measurement apparatus, such as a thermometer, and it takes time for the apparatus, which is of macroscopic dimensions, to react to the system and reach an equilibrium state.  No state which changes rapidly, at a molecular scale, can be observed by the human eye, so we always model a measurement as an infinite time average and define the following notation:
$$\langle f \rangle_t = \lim_{T\rightarrow\infty}\frac1T\int _0^T f(p(t),q(t ))dt .$$
The point is that (p(0),q(0 )) itself is unknown and uncontrollable.  Hence time averages are impossible to calculate, and yet they are what can be measured scientifically. 

On the other hand, if we take $dpdq$ to be Liouville measure on $M$, then phase averages
(we introduce two different notations for this same concept in the following equation)
$$\overline f = \langle f \rangle = \int_\Omega f(p,q)d\omega$$
can be calculated, at least approximately.  Here, $\Omega$ is a compact surface of constant energy within $M$ and the measure is the appropriate invariant measure inherited from Liouville measure.

A dynamical system is said to be ergodic if for all measurable $f$, 
we have
$$\langle f \rangle_t = \langle f \rangle$$
for almost all initial conditions $(p(0),q(0))$ (note that the left hand side depends implicitly on a choice of initial conditions but the right hand side depends only on $f$).

The importance of ergodicity is that if a dynamical system is ergodic, then
macroscopic measurements, the only ones we can make, are reliable guides to 
the phase averages, the only quantities we can really calculate.  Without
ergodicity, there is no way to connect theory with experiment.

If a time series is ergodic then we can use the sample mean to estimate the 
mean, and also use the correlogram to estimate the theoretical auto-correlation function, which then by L\'evy's philosophy tells us everything of interest about the stochastic process.

Linear systems are the opposite of ergodic.  In fact, very few physical systems are known to be ergodic.  In the 60's, Sinai proved that a system of an ideal billiard ball was ergodic.
In 1941, Oxtoby and Ulam proved that `most' dynamical systems are ergodic.
Nevertheless, there is no proof that, e.g., the dynamical system of the weather or coin tossing is ergodic.

Khintchine [9] in 1943 proved that if $R(\tau)\rightarrow 0$ as $\tau\rightarrow\infty$, then $f$ is ergodic, meaning that 
the above equation holds for almost all initial conditions.  But $R$ is not $\varphi$, and in particular it can not be observed directly and it depends on the choice of $\mu$ the probability measure.
For linear systems, $R$ is quasi-periodic and so is $\varphi$.

Statistical Mechanics considers dynamical systems in which the number of degrees of freedom is very large.  Ocean waves can either be modelled by a non-linear wave equation such as Navier--Stokes, with a small number of degrees of freedom, or by a linear Hamiltonian mechanics, at the molecular level, with an astronomical number of degrees of freedom.  Even better, it considers a family of `similar' dynamical systems parametrised by the number of degrees of freedom and considers various limits as $n\rightarrow\infty$.
These limits include the traditional `thermodynamic limit' but as Balian and others have argued, one can define many different types of such limits.

Khintchine observed that since the number of degrees of freedom is very large, asymptotic formulae for the quantities of interest should be obtainable by means of the methods of probability theory, especially its limit theorems.  In particular, he observed that ergodicity in itself was not of central importance since it was asking for too much to have exact equalities of time averages and phase averages for all measurable functions.  It would suffice
to have relations for some physically significant observables which hold asymptotically as the number of degrees of freedom goes to infinity.  In 1943 he published vague but profound and visionary conjectures in this regard, but was unable to establish them in more than a few special cases and even then only with the help of the assumption that 0 = 1, for which he has been much criticised.

\centerline{\bf Khintchine's conjectures}

For a family of dynamical systems $M_n$, parametrised by their (increasing) number of degrees of freedom, and representing in some sense `the same physics', and for certain physically significant quantities, each one represented by an $f_n$, an observable for each $M_n$, again in some sense being `the same' as $n$ increases, Khintchine conjectured that $f_n$ would become approximately ergodic for $n$ sufficiently large.  Ruelle [13] and Lanford [8] were able to make some progress on this for systems with weak and short-range interactions and for observables  that were some sort of average over the entire system, similar to the thermodynamic quantities such as temperature.  Yet Brownian motion is a very well-known
ergodic stochastic process which does not at all fit into this framework: the momentum of one particle becomes, as the number of other particles increases without bound, 
a stationary stochastic process closely related to Brownian motion, known as the Ornstein--Uhlenbeck process.  Since it is the momentum of only one particle, it is not a thermodynamic quantity nor do the methods of Khintchine--Ruelle--Lanford apply.

\centerline{\bf The Gibbs Program and Brownian Motion}

In 1965 Ford--Kac--Mazur [2] showed how Brownian motion could arise in the limit of a sequence of explicit Hamiltonian systems.  Their procedure was a model of carrying out the program envisioned by Willard Gibbs as long ago as 1900.
The breakthrough was to allow a very violent, long-range interaction between the particles, one so violent as to require a kind of renormalisation in the limit.
(In 1961 Schwinger [14] published a very interesting quantum precursor of this. 
Indeed, Schwinger's set-up involved a negative temperature amplifier which amplified quantum motion into a classical stochastic process.)  This successfully carried out Gibbs's program for statistical mechanics for this concrete example, certainly one of central importance, and all the more striking since each system was linear but the limit stochastic process was ergodic.
But they imposed a probability distribution (as did Gibbs himself, and in this respect was criticised by Khintchine), that of Maxwell--Boltzmann, on the dynamical systems by \it fiat\rm, and did not address Khintchine's conjectures.

\vskip -8pt
Their results ought to be robust in the choice of probability distribution and the choice of interaction.  
Students of Kac, Kim [11], and others have pursued this question of robustness.

The key result of Ford--Kac--Mazur is as follows: fix a temperature $T$.  Put the corresponding Maxwell--Boltzmann probability distribution on the space of initial conditions $\Bbb
R ^{2n+2}$.
Then there exists a family of matrices $A_n$, each one giving a linearly coupled system of harmonic oscillators with Hamiltonian $H_n$, such that, with the appropriate cut-off in the interaction to avoid singularities, 
$$R(\tau) \longrightarrow e^{-d\vert\tau\vert}$$
as $n\rightarrow \infty$
where $R$ is the theoretical (phase) auto-correlation function of $p_o$, the momentum of the zero$^{\text th}$ particle (both of which depend on $n$), and $d$ is a constant.
Since one knows the limit of the theoretical auto-correlation functions, then, by
L\'evy's philosophy, one knows which stochastic process (up to equivalence)
ought to be considered the limit of these processes.  This is all the more striking since the 
coupling constants, the entries of the matrices $A_n$, do not possess a limit, but instead grow without bound.

Yet for any finite $n$, $p_o(t)$ is quasi-periodic:
$$p_o(t) = \sum a_n \cos \lambda_{n,k}  t + \sum b_n \sin \lambda_{n,k} t$$
where the $\lambda_{n,k}$ depend on $A_n$.  Hence so is $R_n$ and so is $\varphi_n$.
Hence $R_n\not\longrightarrow 0 $ as $\tau \rightarrow \infty$ for any $n$.  We cannot interchange the limits in $n$ and $\tau$.

\centerline {\bf A sequence of dynamical systems}

For future use, we introduce some common notation.  For $f$ any function on 
the phase space $\Omega$ of a dynamical system, let $f_t$ denote the function composed with the
flow on the system for $t$ units of time.  Let $\langle f\rangle = \int_\Omega 
f(\omega) d\mu$ whatever the invariant measure $d\mu$.
Let $\langle f \rangle_t = \lim_{T\rightarrow \infty} {1\over T} \int _{0}
^T f_t (\omega) dt $ which implicitly depends on $\omega\in\Omega$ although 
this will usually be suppressed in the notation.  
The point is to investigate when $\langle f \rangle_t = \langle f \rangle$
approximately for almost all $\omega$ or at least
the overwhelming majority of $\omega$.


Ford--Kac--Mazur introduced a stylised model [2] of Brownian motion, which 
consists of a mote whose canonical co-ordinates are $p_o$, 
$q_o$, and $n$ particles of the same mass.
We are going to assume $n=2N$ is even and let all vector and matrix 
indices run from $-N$ to $N$.  
  The canonical co-ordinates of the 
$i^{\text{th}}$ harmonic oscillator are $p_i$, $q_i.$  \ \ Let the Hamiltonian 
of this system be $H_n$ and write 
$$H_n=  \sum_{i=-N}^N {p_i^2\over 2m} + 
{1\over2}  (q_{-N}, q_{-N+1}, \dots  q_N) A 
 \pmatrix  q_{-N} \\
 q_{-N+1}\\ \vdots  \\ q_N \endpmatrix$$
where A is a symmetric $n + 1-$square real matrix with positive eigenvalues.

The trajectories of this flow satisfy
(where $A$ depend implicitly on $n$.)  
$$p(t)=\cos (A^{1/2}t)\cdot p(0) -A^{1/2}\sin (A^{1/2}t)\cdot q(0).$$
We focus on $p_0(t)$.
If the particles are all alike, it is 
natural to assume the matrix $A$ is what is 
called, `cyclic'.  Each row is the previous one shifted over by one. 
The eigenvalues of $A$, $\omega^2_i$, satisfy
$$(A)_{ml}={1\over n+1}\sum_{-N}^{N}\omega_i^2e^{{2\pi {\sqrt -1} \over n+1} i(m-l)}$$
where $i\neq \sqrt {-1}$.  
This is obviously symmetric if we make a simple assumption on the $\omega^2_i$'s.

Let $\zeta = e^{i\pi\over 2N+1}$.   
It is a classical fact about cyclic matrices that 
$$(\cos A^{1\over2}\tau )_{mn}={1\over n+1} \sum_{i={-N}}^N \cos (\omega_i \tau)
\zeta^{i(m-n)}
,$$
This formula is an expression of the fact that the vectors (there are $2N+1$
of them as $i$ runs from $-N$ to $N$) $\frac 1 {\sqrt{2N+1}}
(\zeta^{ij})_{j=-N,\dots N}$ are a normal basis of eigenvectors of $A$ with 
eigenvalues $\omega_i^2$.

This formula holds for 
$A^{1\over 2} \sin A^{1\over 2} t$ as well
.
(We omit the proofs of facts about cyclic matrices, which may be found in Gerhard Kowalewski, {\it 
Determinantentheorie} (as cited in Ford--Kac--Mazur [2]).  Hence (here and elsewhere, all indices run from $-N$ to $N$)
$$p_o(t)={1\over n+1}
\left\{\sum_k \sum_i \cos (\omega_it )\zeta^{-ik}p_k(0) 
-\sum_k \sum_i \omega_i\sin (\omega_it )\zeta^{-ik}q_k(0) 
\right\}.$$

We put $\widehat p (k) = \sum_i \zeta^{-ik} p_i(0)$ and similarly for $\widehat q$ and 
rewrite the sums above as
$$ p_o(t) = {1\over n+1} \left\lbrace
\sum_k \widehat p(k) \cos (\omega_k t)
- \sum_k \widehat q(k) \omega_k \sin (\omega_k t)\right\rbrace.$$

  Define  the auto-correlation (sometimes called the auto-covariance) function of 
this trajectory by 
$\varphi_n(\tau)
= \langle p_o(t)p_o(t+\tau)\rangle_t.$
For each $\tau$, $\varphi_n(\tau)$ is a physical observable on $M_n$.  (As $\tau$ varies,
we have a uniform family of physical observables.)  Just as all the $M_n$ have, 
intuitively, the same physical meaning, so too these observables 
(or, rather, uniform families of observables) all have the `same  
physical meaning'.  Intuitively, each one measures how `random' the trajectory (through 
a given point of phase space) is.   It is a descriptive statistic of a definite 
set of data: whatever the initial conditions `really are', the data is the future 
path of the trajectory (or the past, it makes no difference), and this is a 
deterministic descriptive statistic.  
The main result of this paper is to make 
rigorous the notion that in the limit as $n\rightarrow\infty$, all normal 
trajectories have the same autocorrelation function, the Markoffian one which 
represents maximal randomness possible in this situation.  

\vskip -8pt
But since this intuitive notion of limit is problematic, we make this notion 
rigorous by talking about normal cells for finite $n$.  For fixed $n$, we can 
define a normal trajectory as being one which (within certain limits of approximation)
has the same auto-correlation function as 
all other normal trajectories, viz., the one which is the 
best possible approximation to the Markoffian exponential decay. 

\vskip -8pt
In general, the physicist Sir James Jeans outlined a common-sensical view of the 
foundations of Statistical Mechanics which has had an influence on his followers, 
Darwin and Fowler, and through them, on Khintchine, but has not penetrated as fully 
into the consciousness of philosophers of science [16] as it deserves (although the 
Ehrenfests consider it carefully in [1]).  
Statistical Mechanics is defined as the study of the statistical properties of the 
normal trajectory.  By statistical is meant, descriptive statistics, so the notion 
of probability does not enter into this definition.  We, following Wiener, will 
mean the auto-correlation function $\langle f \rangle_t= \varphi(t) $ of a trajectory. 
The important thing is to define 
normal.  Jeans [4] defined
a normal property of a state (or trajectory) to be 
a property which is possessed by the overwhelming majority of states in the system, 
so that, as the number of degrees of freedom increases without bound, the states 
which do not possess that property possess negligible Liouville measure. 
A state is then defined as a normal state if it possesses all those normal properties 
`of which it is capable.'  
This definition of normal
 was not given with full logical rigour: the task of this paper is to fix that in an important example.


\centerline {\bf Some random finite trigonometric sums}

\vskip -8pt
Normalise the measure on the surface of constant energy, $\Cal M _E$, inherited from 
Liouville measure to be total mass unity.
In this paper the only properties we are concerned with are  $\varphi (t)$.  That 
is, a normal cell is a sequence of subsets $\Cal N_n$ of $\Cal (M_n)_{E_n}$ such that:
for every choice of three positive epsilons, we have for $n$ sufficiently large that 
 it has measure $1-\epsilon$ 
and the $\varphi (t)$ are within $\epsilon_1$ of each other for $t< 1/\epsilon_2$.
The energy level $E_n$ is defined for traditional reasons, and to make the 
comparison with traditional results convenient, to be that energy level which 
is most probable according to the Maxwell distribution: it is $n+1\over kT$.

Intuitively, this would mean that the limits of the auto-correlation functions 
(uniform convergence on compact sets) are the same for points in the same cell.

The existence of a normal cell is the kind of approximate ergodicity 
analogous to what Khintchine envisioned, (as is, in a very different way, the dispersion theorem proved by Khintchine and Lanford). 
It is well-known that the limit stochastic process Ford--Kac--Mazur 
constructed from this sequence is ergodic, since it is the Ornstein--Uhlenbeck process.

It is elementary that in general all for any trigonometric 
sum
$$p_o(t) = \sum_i a_i \cos ( \omega_i t) + b_i \sin (\omega_i t),$$
(here, we may and do assume all $ \omega_i>0$) the auto-correlation function is 
$$\varphi (\tau) = \sum_i \frac12 (\vert a_i\vert^2 + \vert b_i \vert^2)
\cos ( \omega_i t) .$$ 
Applying this result to our formula for $p_o$, 
we obtain 
$$\varphi (\tau) = \sum_k \frac12 \left({1 \over 2N+1}\right)^2
(\vert \widehat p(k) \vert^2 +
 \vert \omega_k \widehat q(k) \vert^2)
\cos ( \omega_k t) .$$ 
In order to show that normal cells exist, we want to show that 
the measure of initial conditions in phase space which yield 
approximately the same $\varphi(\tau)$ tends to unity as $n\rightarrow
\infty$.  We may regard this as a random trigonometric sum.
(Regarding the $p_i(0) $ and the $q_i(0)$ as random variables).
We show that the variance of $\varphi(\tau)$ is 
negligible for large $n$.  
In 1866  
(see Stroock [15] p.\ 77)
Mehler proved that for 
$x_1^2 + x_2^2 + \dots x_n^2 = \rho n$ and   
the uniform surface area measure on the surface of this sphere
of total mass one,
then as $n\rightarrow
\infty$, $x_1$ tends weakly to a Gaussian random variable with 
mean zero and standard deviation $\rho$.  
 The rate of convergence can be controlled 
explicitly, this is merely a concrete calculation of surface areas on 
spheres.  

It is obvious that the coordinates are uncorrelated, that 
each $x_i$ is perfectly un-correlated with the $x_j^2$ for 
$j\neq i$, and that the squares are negatively correlated 
with each other.  Hence $Var ( x_i^2)$ is approximately two,
and $Var (x_i^2 + x_j^2) \leq Var(x_i^2) +   Var(x_j^2)$.    

We wish to estimate the variances of the $\widehat p(k) $ and the
$\omega_k\widehat q(k)$. 

Because $H(p,q) = H(\widehat p,\widehat q) $ and, more precisely,
because $$ (q_{-N}(0) , \dots q_{N}(0)) \cdot A 
 \cdot \pmatrix  q_{-N}(0) \\
 q_{-N+1}(0)\\ \vdots  \\ q_N(0) \endpmatrix
= (\widehat q(k) , \dots \widehat q(k))(\zeta^{-ml}) \cdot A\cdot (\zeta^{lm}) 
 \pmatrix  \widehat q(k) \\
 \widehat q(k)\\ \vdots  \\ \widehat  q(k) \endpmatrix
$$ (this is because the matrix $(\zeta^{ml})_{lm}$ is the change 
of basis matrix that diagonalises $A$) 
it follows that$$ (q_{-N}(0) , \dots q_{N}(0)) \cdot A 
 \cdot \pmatrix  q_{-N}(0) \\
 q_{-N+1}(0)\\ \vdots  \\ q_N(0) \endpmatrix
= \sum_k \widehat q(k)^2 \omega_k^2.$$

But 
$2H - \sum_k \widehat q(k)^2 \omega_k^2 = \sum_i  p_i(0)^2 .$
Hence $\sum_k \widehat q(k)^2 \omega_k^2$ and $\sum_i  p_i(0)^2 $
are perfectly anti-correlated and hence have equal variances.
But this last is bounded by 2(2N+1).
Now the $\omega_k$ are all real and positive in the applications we have 
in mind for later.  Also, since the matrix is symmetric, $\omega_{-k} 
= \omega_k$.  
Hence, obviously, we may arrange that the $\widehat q(k)$ are real without 
altering the auto-correlation function (since $\widehat q(k) = \overline{
\widehat q(-k)}$).
Hence$$Var{1\over 2N+1}{\frac12\over 2N+1} \sum_k \vert \omega_k \widehat q(k) | ^2 \cos 
(\omega_k \tau)
< {1\over(2N+1)^2}{\frac14\over (2N+1)^2} 2(2N+1).$$ 

Since, by definition, $\widehat p (k) = \sum_i \zeta^{-ik} p_i(0)$, 
we have that $\widehat p (k) \over \sqrt{2N+1}$ is related to $p_i(0)$ by a 
unitary transformation, the sum of squares of the moduli of the 
coordinates does not change.  Then neither does their variance.  Hence 
the variance of $\sum\frac12\widehat p(k)\over (2N+1)^2 $
is bounded by ${2(2N+1) \frac14\over (2N+1)^2} = {\frac12\over (2N+1)}$.

This very weak ergodicity does not depend on any properties of 
the $\omega_k$ except that the matrix $A$ is cyclic and symmetric.  It holds 
even when the dynamics does not tend towards Brownian motion
(which only happens for a very specific choice of $\omega_k$).

We wish to relate Wiener's time auto-correlation function, a deterministic 
concept, 
to the phase auto-correlation function, at least when the trajectory 
is normal.  Note first that 
as usual, the Maxwell distribution `bunches up' for very 
large $n$ around the most probable energy value, and thus the average 
taken with respect to the Maxwell distribution over the entire, unbounded, 
phase space is the same as the average over one energy level ellipsoid, 
with respect to the uniform distribution.  An elementary part of what
Khintchine proved [9] p.\ 68, is

\noindent \bf Theorem\rm\ (Khintchine):
         Let $d\mu_E$ be the normalised measure on the constant energy 
shell $\Omega_E$ of $\Cal M _n$ with energy level $H=E$ inherited from 
Liouville measure.  Then $d\mu_E$ is invariant 
under the flow.  
Let $\varphi^\omega(\tau)$ be Wiener's time-autocorrelation function 
for a given initial condition $\omega\in\Omega_E$.
Then the expectation of the time-autocorrelation function is equal to the 
phase auto-correlation function, \it i.e.\rm, 
$$\int_{\Omega_E}  \varphi^\omega(\tau) d\mu(\omega) = 
\int_{\Omega_E} f_t(\omega)f_{t+\tau}(\omega) d\mu(\omega). $$ 

\centerline {\bf Time auto-correlation functions of the Ornstein--Uhlenbeck process}

We now specify a precise dynamics by explicitly choosing the $\omega_k$.  
Let $\omega_s=\tan {\pi s \over n+1}$.  

Ford--Kac--Mazur calculate the phase auto-correlation of each finite stage and 
pass to the limit (with a cut-off renormalisation) obtaining the usual 
auto-correlation function of the Ornstein-Uhlenbeck process, 
$\pi e^{-|\tau|}$.  

It is part of what they prove that
  for any compact set $K = \{\tau \in [0,K] \}$, there exists an $N$ so 
large that 
$$
\int_{\Omega_E} p_o(0)(p_o)_\tau d\mu(\omega)   $$ 
approximates to $\pi e^{-|\tau|}$ to any desired accuracy.

Their method of proof was relatively elementary and can be included here.
     A simple trick changes this into a standard cosine transform 
which can be looked up in any table of integrals.  Let $u=\tan\theta$.  
Then $\theta=\arctan u.$
This integral is, then, equal to one of the Riemann sums for the 
improper integral 
$$\int_{-\infty}^\infty
\cos(\tau u){du\over u^2+1}$$
This, as aimed at, is the cosine transform of the bump function 
$1\over u^2+1$.  It is equal to $\pi e^{-\tau}$
when $\tau$ is positive, but it is symmetric since cosine is an even function, 
so it is equal to 
$\pi e^{-|\tau|}$.  
There is no problem with convergence in this calculation as we let 
$\epsilon \rightarrow 0$; the improper integral is very nicely behaved.


\noindent\bf Theorem: \rm Suppose given $\epsilon, \delta > 0$ and $K$.  Then there exists an 
$N$ so large that the measure of the set of trajectories such that $\vert \varphi(\tau) - \pi e^{-|\tau|} \vert
> \epsilon$ for all $\tau\in K$ is less than $\delta$.  (This implies that a normal cell exists, the 
cell of all trajectories such that on $K$, their auto-correlation functions are 
within $\epsilon$ of the phase auto-correlation function $\langle f_of_\tau\rangle$ .)

\noindent \bf Proof\rm.  \ \ Since the variance of $\varphi(\tau)$ for any fixed $\tau$ is less than $1\over(2N+1)$,
Tchebycheff's inequality gives us that the measure of the set of trajectories such that $\varphi(\tau)$ 
differs by more than $\epsilon$ from 
$\langle f_of_\tau\rangle$, which is its expected value, is less than 
$ 1\over \epsilon n$.  If we cover the interval $K$ with a uniform mesh of width $\Delta x$
then there are $K\over\Delta x$ values of $\tau_i$ in this mesh.  If we treat the 
measure of the set of trajectories yielding  a deviation of $\varphi(\tau_i)$ by more than $\epsilon$ from its
expectation for each $i$ as independent events, which they are not, then the 
measure of the set of initial conditions such that even one of these violations will occur is less than 
$K\over n \epsilon\Delta x$.  

\vskip -8pt
As $n$ grows, the difference between the quasi-periodic expectation of 
$\varphi(\tau)$ and its limit the exponential decay $\pi e^{-|\tau|}$, may be 
arranged to be less than $\epsilon$ on any compact set, especially $K$.  And this latter function clearly satisfies 
$| \Delta y | < \Delta x$ over this or any other mesh.  
Hence, if $\tau$ varies over a region of width $\Delta x$, 
$\varphi(\tau)$ varies by less than $2\epsilon + \Delta x$.  
We may take $\Delta x = .5\epsilon$.  
and $n> K\frac2{\epsilon^2}$ to get a 4$\epsilon$ proof.  
\it Q.E.D.\rm\

This method yields an essentially independent proof of the results of Ford--Kac--Mazur and their generalisations.
It does not seem as if the usual methods of large-sample theory of time series, or of Khintchine, Ruelle, or Lanford, or the usual limit theorems of probability theory, can be used to obtain this or similar results.
I would like to conjecture that normal cells in this sense exist for a much wider class of sequences of 
dynamical systems.

\vskip -8pt
Now for all practical purposes, a stochastic process can be replaced by its 
auto-correlation function.  In fact, a Gaussian stationary stochastic process 
is determined up to equivalence by the phase auto-correlation function.  
As $n$ increases without bound, the time auto-correlation functions of normal 
trajectories approaches a limiting function.  We may define a stochastic process
by the requirement that its
phase auto-correlation function be this limiting function.   Thus we have 
defined the thermodynamic limit of this sequence of dynamical systems as a 
stochastic process.  The sample space of this process has nothing to do with any Hamiltonian 
dynamical system or Liouville measure.  There does not seem to me to be any point in trying
to define a new class of dynamical system, with an infinite number of degrees of freedom,
which would be the limit object here: this would go against L\'evy's philosophy, and 
would be subject to the use of Occam's razor.

\vskip -8pt
In [5] and [6], I have shown how a quantum analogue at negative temperature, 
which is much simpler than the classical case, has many of the same features 
as the model of this paper.  It would be important to generalise the results 
of this paper to a negative temperature heat bath around the mote.
Schwinger in [14] has treated the case of quantum negative temperature 
Brownian motion, claimed it acts as an amplifier (which is understandable, 
since it is done in [5]), and claimed that it amplifies quantum motion to the 
classical level.  The derivation lacks rigour and uses the usual imprecise 
notions of probability.  This is an important topic for the future.

\noindent\bf ACKNOWLEDGMENTS
\rm I would like to thank the Canadian NSERC for partial financial support during
this research and the Dept. of Mathematics and Statistics at Queen's University, Canada,
for its hospitality in 2007--2009, where this research was begun.

\centerline{\bf References}

\noindent [1] P. Ehrenfest and T. Ehrenfest, \it The Conceptual Foundations of the Statistical Approach in Mechanics, \rm Leipzig, 1912, \S 18b, p.\ 41.

\noindent [2] G. Ford, M. Kac, and P. Mazur, `Statistical Mechanics of Assemblies of Coupled Oscillators', \it J.\ Math.\ Phys.\ {\bf6} \rm(1965), 504.

\noindent [3] W. Fuller, \it Introduction to Statistical Time Series, 2nd Ed.\rm, Wiley, New York, 1996. 

\noindent [4] J. Jeans, \it Dynamical Theory of Gases\rm, Cambridge, 1904, III \S\S 52,58 and V \S 87. 

\noindent [5] J. Johnson, ``Statistical Mechanics of Amplifying Apparatus'', to appear, 
 \it Proc.\ VIII International Wigner Symposium, New York City, 2003 \rm eds.\ 
S.\ Catto and B. Nicolescu.

\noindent [6] J. Johnson, ``Thermodynamic Limits, Non-commutative Probability, and Quantum entanglement'', in {\it Quantum Theory and Symmetries, Proceedings of the Third International Symposium,} Cincinnati, 2003, edited by P. Argyres \it et al\rm., Singapore, 2004, 133.

\noindent [7] J. Johnson,  ``Probability as a Multi-Scale Phenomenon'', submitted. 

\noindent [8] O. Lanford, `Entropy and Equilibrium States in Classical Statistical Mechanics', in A. Lenard (ed.), \it Statistical Mechanics and Mathematical Problems, Springer Lecture Notes in Physics\rm, Springer Verlag, Berlin, 1973. 

\noindent [9] A. Khintchine, {\it Matematichiskie Osnovaniya Statisticheskoi Mekhaniki}, Moscow, 1943, English translation, {\it Mathematical Foundations of Statistical Mechanics}, New York, 1949. 

\noindent [10] A. Khinchin, {\it Ob Analiticheskom Apparate Fizicheskoi Statistiki}, Moscow, 1950, English translation, {\it Analytical Foundations of Physical Statistics}\rm, Delhi, 1961.

\noindent [11] S.\ Kim, ``Brownian Motion in Assemblies of Coupled Harmonic Oscillators''\it J.\ Math.\ Phys., \bf 15 \rm(1974), 578.


\noindent [12] P. L\'evy, \it Processes Stochastiques et Mouvement Brownien, \rm Paris, 1948. 

\noindent [13] D. Ruelle, \it Statistical Mechanics:  Rigorous Results\rm, New York, 1969.

 \noindent [14] J. Schwinger, ``Brownian Motion of a Quantum Oscillator'', \it J. Math.\ Phys.\  \rm {\bf2} no.\ 3 (1961), 407.

\noindent [15] D. Stroock, \it Probability Theory, \rm Cambridge, 1993.

\noindent [16] J. von Plato, ``Ergodic Theory and the Foundations of Probability,'' in {\it Causation, Chance, and Credence, Proceedings of the Irvine Conference on Probability and Causation}, edited by B. Skyrms and W. Harper, vol.\ 1, Dordrecht, 1988, pp.\ 257--277.
\end